\documentclass[aps, prd, twocolumn, nofootinbib, showpacs]{revtex4-2}

\usepackage{amsmath, graphicx, amssymb, dcolumn}
\usepackage{bm, accents, mathrsfs, comment, subfigure}
\usepackage[colorlinks, linkcolor=blue, citecolor=blue, urlcolor=blue]{hyperref}

\def\prl{Phys. Rev. Lett.}
\def\prd{Phys. Rev. D}
\def\aap{Astron. Astrophys.}
\def\apjl{Astrophys. J. Lett.}
\def\apj{Astrophys. J.}
\def\jcap{JCAP}
\def\mnras{Mon. Not. Roy. Astron. Soc.}
\def\sovast{Soviet Ast.}

\newcommand{\ep}{\epsilon}
\newcommand{\Mpl}{M_{\scriptscriptstyle\rm  Pl}}
\newcommand{\ph}{\phi_{\scriptscriptstyle\infty}}
\newcommand{\W}{\scriptscriptstyle\rm WD}
\newcommand{\N}{\scriptscriptstyle\rm NS}
\newcommand{\sun}{\scriptscriptstyle\odot}
\newcommand{\fR}{f^{\prime}\!(R_{\scriptscriptstyle\infty}\!)}
\newcommand{\gr}{\scriptscriptstyle\rm GR}
\newcommand{\ac}{\accentset}

\begin{document}

\title{Tests of gravitational scalar polarization and constraints of chameleon $f(R)$ gravity from comprehensive analysis of binary pulsars}
\author{Xing Zhang$^{1,2}$}\email{zhxing@nwu.edu.cn}
\affiliation{$^1$School of Physics, Northwest University, Xi’an 710127, China}
\affiliation{$^2$Shaanxi Key Laboratory for Theoretical Physics Frontiers, Xi’an 710127, China}

\begin{abstract}
Chameleon $f(R)$ gravity is equivalent to a class of scalar-tensor theories of gravity with chameleon screening mechanism allowing the theory to satisfy local tests of gravity.
Within the framework of chameleon $f(R)$, we study the impact of the chameleon mechanism on the orbital evolution of binary pulsars, and calculate in detail the post-Keplerian (PK) effects (periastron advance, Einstein delay, Shapiro delay, orbital period decay and eccentricity decay) of binary orbit.
The differences in PK effects between general relativity (GR) and chameleon $f(R)$ are elegantly quantified by a combination of star's compactness and theory parameter.
We use the mass-radius relation to break the degeneracy between these two parameters, thus allowing us to constrain the theory.
We simulate the temporal evolution of the orbital period and eccentricity of neutron star (NS) - white dwarf (WD) binaries, and the results indicate that the orbital evolution is typically faster than in GR due to the emission of dipole radiation in chameleon $f(R)$.
We use the observables of PK parameters from the three NS-WD binary pulsars to place constraints on chameleon $f(R)$ and possible deviations from GR by performing Monte-Carlo simulations.
We find that PSR J1738$+$0333 is the most constraining test of chameleon $f(R)$ in these systems.
Our results show no solid evidence of the existence of helicity-0 or helicity-1 polarization states inducing dipole radiation, exclude significant strong-field deviations and confirm that GR is still valid for strong-field asymmetric systems.
\end{abstract}

\maketitle

\section{Introduction}\label{section1}
Although Einstein's theory of general relativity (GR) is indeed the most successful theory of gravity, it suffers from the quantization \cite{Kiefer:2007aa, DeWitt:1967yk} as well as dark matter and dark energy problems \cite{Cline:2013aa, Sahni:2004ai}. 
Therefore, testing GR still is one of the key tasks in modern physics \cite{Will:1993aa}. 
Studies of alternative theories of gravity play a significant role in testing GR.

A natural alternative theory is $f(R)$ gravity \cite{Sotiriou:2010aa, De-Felice:2010aa}, in which the Ricci scalar in the Einstein--Hilbert action is replaced by a general function of the Ricci scalar. 
The $f(R)$ theories do not seem to introduce any new type of matter and can drive early inflation \cite{Starobinsky:1980aa} or late-time acceleration of the universe \cite{Capozziello:2003tk, Carroll:2004aa}.
In fact, the $f(R)$ theories can be reformulated in terms of scalar-tensor theories with a strong coupling of the scalar field to matter \cite{Sotiriou:2010aa, De-Felice:2010aa}. 
The strong coupling would induce the scalar fifth force in the theory, which violates all current experimental constraints on deviations from Newton's law of gravity.  
In order to evade these tight local tests of gravity, the chameleon mechanism \cite{Khoury:2004aa, Khoury:2004ab, Gubser:2004aa} is introduced into $f(R)$ theories, which imposes restrictions on the functional form of $f(R)$.
The chameleon scalar field can develop an environment-dependent mass, which increases as the ambient density increases.  
Therefore, the scalar fifth force can be hidden and evade the tight local tests in high density regions (e.g., the solar system), in which the force range becomes so short that it is extremely difficult to detect by local test experiments \cite{Gubser:2004aa}. 
Whereas in low density regions (e.g., the galaxy or the universe), the scalar fifth force becomes the long-range force, which could affect the galactic dynamics \cite{Gronke:2015aa, Schmidt:2010aa} and the evolution of the universe \cite{Starobinsky:1980aa, Capozziello:2003tk, Carroll:2004aa}.

Since the Hulse-Taylor binary pulsar observations led to the first indirect detection of gravitational waves (GWs) \cite{Hulse:1975aa, Taylor:1982aa, Taylor:1989aa}, binary pulsars have become the excellent laboratories for testing gravity in the strong field regime \cite{Stairs:2003aa, Damour:1996aa, Damour:1998aa, Yagi:2014aa, Beltran-Jimenez:2016aa}.
In this paper, we study the full post-Keplerian (PK) effects of binary pulsars in the framework of chameleon $f(R)$.
We calculate the effects of periastron advance, Einstein delay and Shapiro delay by investigating the orbital dynamics of binary pulsars, and derive the decay rates of orbital period and eccentricity caused by GWs damping by investigating the Noether charges and currents in the theory.
In chameleon $f(R)$, the leading term of tensor GWs radiation is the quadrupole radiation carrying both energy and angular momentum, and the leading term of scalar GWs radiation is the monopole radiation carrying energy but not angular momentum. 
However, the monopole radiation and the quadrupole radiation are of the same post-Newtonian (PN) order.
Scalar dipole radiation carries both energy and angular momentum away from the binary pulsars and dominates the orbital decay, and its intensity is proportional to the square of the difference in the compactnesses of binaries.
Therefore, the asymmetric systems like the neutron star (NS) - white dwarf (WD) binary pulsars are the ideal targets for testing chameleon $f(R)$ gravity. 
We perform the numerical simulation of the orbital evolution of binary pulsars, and place constraints on chameleon $f(R)$ with the observables of PK parameters from three NS-WD PSRs J1141$-$6545, J1738$+$0333 and J0348$+$0432.
It turns out that the dipole radiation in chameleon $f(R)$ further accelerates the orbital evolution of binary pulsars.
The orbital period decay rates from these three systems impose the tight constraints on dipole radiation, which can also be thought of as tests of the existence of helicity-0 or helicity-1 degrees of freedom.
The pulsar constraint from PSRs J1738$+$0333 is the most stringent test in these three systems.
These pulsar tests rule out the significant deviations from GR in strong-field asymmetric systems.

The organization of this paper is as follows. 
In Sec. \ref{section2}, we review $f(R)$ gravity and chameleon mechanism.
In Sec. \ref{section3}, we calculate in detail the PK parameters in chameleon $f(R)$ gravity. 
In Sec. \ref{section4}, we place constraints on chameleon $f(R)$ by the observational data of the binary pulsar, and discuss in detail these results. 
We conclude in Sec. \ref{section5}.
Appendixes present further mathematical details.

\section{$f(R)$ gravity with chameleon screening mechanism}\label{section2}
The $f(R)$ gravity is based on the corrections and extensions of GR adding higher order terms or non-minimally coupled scalar fields into the dynamics.
The Lagrangian density for $f(R)$ gravity takes the form \cite{Sotiriou:2010aa, De-Felice:2010aa}
\begin{eqnarray}
\label{Lagrangian_fr}
\mathcal{L}=\frac{\Mpl^2}{2}\sqrt{-g} f(R)+ \mathcal{L}_m(g_{\mu\nu},\psi_m),
\end{eqnarray}
where $\Mpl\equiv\sqrt{1/8\pi G}$, $G$ is the gravitational constant, $g$ is the determinant of the metric $g_{\mu\nu}$, $R$ is the Ricci scalar, $\mathcal{L}_m$ is the matter Lagrangian density, and $\psi_m$ denotes all the matter fields. Here, we set the units to $c=\hbar=1$.
The $f(R)$ gravity can be recast as a scalar-tensor theory via the following conformal transformation \cite{Sotiriou:2010aa, De-Felice:2010aa}
\begin{eqnarray}
\label{conf_fR_ST}
g_{\mu\nu}\rightarrow \tilde{g}_{\mu\nu} = g_{\mu\nu} f'(R) \equiv g_{\mu\nu}\exp(-\sqrt{\frac{2}{3}}\frac{\phi}{\Mpl}),
\end{eqnarray}
where a prime denotes differentiation with respect to $R$, and $\phi$ is the scalar field which can be directly related to the Jordan frame Ricci scalar by the above relation.
The Lagrangian density in the Einstein frame has the form \cite{Sotiriou:2010aa, De-Felice:2010aa}
\begin{eqnarray}
\label{Lagrangian_fR_E}
\tilde{\mathcal{L}} \!=\! \sqrt{\!-\tilde{g}}\Big[\frac{\Mpl^2}{2} \! \tilde{R} \!-\! \frac{(\tilde{\partial}\phi)^2}{2} \!-\! V\!(\phi)\Big]\!+\!\tilde{\mathcal{L}}_m(A^2\!(\phi)\tilde{g}_{\mu\nu},\psi_m),~~~~
\end{eqnarray}
where the potential is
\begin{eqnarray}
V(\phi)=\frac{1}{2}{\Mpl^2}f'(R)^{-2}(f'(R)R-f(R)),
\end{eqnarray}
and the coupling function is 
\begin{eqnarray}
A(\phi)=f'(R)^{-\frac{1}{2}}=\exp(\frac{\phi}{\sqrt{6}\Mpl}).
\end{eqnarray}
Here, a tilde represents quantities in the Einstein frame.

Variation of $\tilde{\mathcal{L}}$ with respect to the tensor field and the scalar field gives the field equations
\begin{eqnarray}
\label{tensor_Eq}
\tilde{G}_{\mu\nu}\!
&=& 8\pi G \big[\tilde{T}_{\mu\nu}+\partial_\mu\phi\partial_\nu\phi-\big((\tilde{\partial}\phi)^2/2+V\big)\tilde{g}_{\mu\nu}\big],~~~~~
\\ 
\label{scalar_Eq}
\tilde{\square}\phi
&=& {d V_{\rm eff}}/{d\phi},
\end{eqnarray}
where $\tilde{\square}$ is the curved space d'Alembertian, $\tilde{G}_{\mu\nu}$ is the Einstein tensor, and $\tilde{T}_{\mu\nu} \equiv(-2/\sqrt{-\tilde{g}}) \delta ({\int} dx^4 \tilde{\mathcal{L}}_m) / \delta \tilde{g}^{\mu\nu}$ is the energy-momentum tensor of the matter.
Here, the effective potential $V_{\text {eff}}(\phi) \equiv V(\phi)+{\rho}A(\phi)$, and $\rho$~\footnote{$\rho$ is defined as the conserved energy density in the Einstein frame \cite{Joyce:2015aa}.} is the local environment density of the scalar field.
In the previous work \cite{Zhang:2016aa, Zhang:2017aa, Zhang:2019ab, Zhang:2019aa, Zhang:2019ac, Liu:2018ab}, we have investigated the screening mechanisms for the Lagrangian density \eqref{Lagrangian_fR_E} with a general potential and coupling function.
For this theory to have a screening mechanism one must require that \cite{Zhang:2019ab}
\begin{eqnarray}
\label{mass_eff}
\frac{d V_{\rm eff}}{d\phi}\bigg|_{\phi_{\rm min}}=0,  ~~  
m^2_{\rm eff}\equiv \frac{d^2 V_{\rm eff}}{d\phi^2}\bigg|_{\phi_{\rm min}}>0,  ~~   
\frac{d m_{\rm eff}}{d\rho}>0.~~~~~
\end{eqnarray}
In other words, the effective potential has a minimum (acting as the physical vacuum), and the effective mass of the scalar field increases as the ambient density increases. 
As a result of these requirements, all mechanics effects induced by the scalar field are suppressed in dense regions, where the range of the scalar fifth force is so short that it is hard to find by local experiments.
Theories in which the scalar field mass depends on the ambient density are called to be chameleon theories \cite{Khoury:2004aa, Khoury:2004ab, Gubser:2004aa}.
For $f(R)$ gravity the above requirements turns into, in some regions of $\phi$ \cite{Brax:2008aa},
\begin{eqnarray}
\label{fR_V_const}
\frac{d V}{d\phi}<0,   \quad~~   
\frac{d^2 V}{d\phi^2}>0,   \quad~~  
\frac{d^3 V}{d\phi^3}<0,
\end{eqnarray}
these can be translated into the constraints on the functional form of $f(R)$ (see Appendix \ref{appendix_cfR}).
If the potential function satisfies the above conditions, the $f(R)$ gravity can have a chameleon screening mechanism. 
The $f(R)$ gravity with chameleon screening mechanism is also called chameleon $f(R)$ gravity.

Note that, for convenience, thereafter, we still use $f(R)$ to refer to chameleon $f(R)$, work in the Einstein frame and no longer label the Einstein frame with a tilde.

\section{PK parameters}\label{section3}
In this section, we study the PK effects in $f(R)$ gravity and calculate in detail the PK parameters for the binary pulsar moving on a quasi-elliptical orbit.

\subsection{Periastron Advance}
The periastron advance is an astronomical phenomenon in which the major axis of the orbit slowly rotates in the orbital plane.
This phenomenon is because in fact the net force experienced by a planet does not vary exactly as inverse-square.

In $f(R)$ gravity, the scalar fifth force modifies the orbital dynamics of binary pulsars and contributes to the periastron advance.
The scalar field corrections to the orbital dynamics can be effectively described by the point-particle action with scalar field-dependent mass introduced by Eardley \cite{Eardley:1975aa}.
The Lagrangian for the $a$-th body is given by
\begin{eqnarray}
\label{matter_Lag}
L_a = m_a(\phi)\frac{d\tau_a}{dt} = m_a(\phi)\Big(-g_{\mu\nu}\frac{dx_a^{\mu}}{dt}\frac{dx_a^{\nu}}{dt}\Big)^{\frac{1}{2}}.~~~~
\end{eqnarray}
By substituting the post-Newtonian (PN) expressions of the scalar and tensor fields in Eqs.\eqref{PN_metric_scalar}, and adopting the method of Einstein, Infeld and Hoffmann \cite{Einstein:1938aa}, we obtain the $N$-body Lagrangian up to $\mathcal{O}(v^4)$,
\begin{subequations}
\begin{eqnarray}
L_{N} 
&=& - \sum_am_a \big( 1-\frac{v_a^2}{2}-\frac{v_a^4}{8} \big) + \frac12\sum_{a}\sum_{b\ne a}\frac{Gm_am_b}{r_{ab}}
\nonumber\\
&& \times \Big[ \mathscr{G}_{ab} + 3\mathscr{B}_{ab}v_a^2 - \frac12( \mathscr{G}_{ab}+6\mathscr{B}_{ab} )( \mathbf{v}_a\cdot \mathbf{v}_b )
\nonumber\\
&& - \frac12\mathscr{G}_{ab}(\mathbf{n}_{ab} \cdot  \mathbf{v}_a)(\mathbf{n}_{ab} \cdot  \mathbf{v}_b) - \sum_{c \ne a} \frac{Gm_c}{r_{ac}}\mathscr{D}_{abc}\Big], \qquad~
\end{eqnarray}
with
\begin{eqnarray}
\mathscr{G}_{\!ab} \!=\! 1 \!+\! \frac{\ep_a\ep_b}{2},~
\mathscr{B}_{\!ab} \!=\! 1 \!-\! \frac{\ep_a\ep_b}{6},~   
\mathscr{D}_{\!abc} \!=\! 1 \!+\! \frac{\ep_a(\ep_b \!+\! \ep_c)}{2},\qquad~
\end{eqnarray}
\end{subequations}
where $\mathbf{n}_{ab}\equiv({\mathbf{r}_a-\mathbf{r}_b})/{r_{ab}}$ is the unit direction vector, and $\ep$ is the scalar charge of the body.
The scalar charge characters the difference from GR and can be well approximated by $\ep=\ph/({\Mpl\Phi})$ (see Eq.~\eqref{epsilon_a}).
Using Eq.\eqref{conf_fR_ST}, the scalar charge can be rewritten as
\begin{eqnarray}
\label{epsilon_fR}
\ep=-\sqrt{\frac{3}{2}}\frac{\ln{\fR}}{\Phi},
\end{eqnarray}
where $R_{\scriptscriptstyle\infty}$ is the background value of Ricci scalar, and $\Phi=Gm/R$ is the compactness of the body and $R$ is its radius.

Specializing to a two-body system (labeled by 1 and 2), the two-body equations of motion following from this Lagrangian are
\begin{eqnarray}
\frac{d^2 \mathbf{r}_1}{dt^2} 
&=& - \frac{Gm_2\mathbf{n}_{12}}{r^2} \Big[ \mathscr{G}(1-v_{1}^2+\frac{\mathbf{v}_{12}^2}{2}-\frac{3}{2}(\mathbf{v}_{2}\cdot\mathbf{n}_{12})^2) \nonumber\\
&& - \frac{Gm_2}{r}(3\mathscr{G}\mathscr{B} \!+\! \mathscr{D}_{122}) \!-\! \frac{Gm_1}{r}(\mathscr{G}^2 \!+\! 3\mathscr{G}\mathscr{B} \!+\! \mathscr{D}_{211}\!) \nonumber\\
&& + \frac{3\mathscr{B}}{2}\mathbf{v}_{12}^2 \Big] \!+\! \frac{Gm_2\mathbf{v}_{12}}{r^2}(\mathscr{G}\mathbf{v}_1 \!+\! 3\mathscr{B}\mathbf{v}_{12}) \!\cdot\! \mathbf{n}_{12},~~~~~~~~\\
\frac{d^2 \mathbf{r}_2}{dt^2} 
&=& \{1 \leftrightarrow 2\}, \nonumber
\end{eqnarray}
where $\mathbf{v}_{12} \equiv \mathbf{v}_1-\mathbf{v}_2$, $r \equiv r_{12}$, $\mathscr{G} \equiv \mathscr{G}_{12}$ and $\mathscr{B} \equiv \mathscr{B}_{12}$. 
Obviously, at the Newtonian order, the equations of motion satisfy the inverse-square law, only the gravitational constant is replaced by $\mathscr{G}G$.
This result also suggests that the conservative orbital dynamics at the Newtonian order still hold, e.g., the Kepler's third law $a^3 = \mathscr{G}{G}m({P_b}/{2\pi})^2$.

Using the above equations of motion, employing the method of osculating elements \cite{Will:1993aa}, the periastron advance of the binary system is given by \cite{Will:1993aa}
\begin{eqnarray}
\label{PK_ome_dot}
\dot{\omega}=\frac{6\pi{Gm}}{a(1-e^2)P_b}\big(\mathscr{B}+\frac{\mathscr{G}}{6}-\frac{m_1\mathscr{D}_{211}+m_2\mathscr{D}_{122}}{6\mathscr{G}m}\big),~~~~
\end{eqnarray}
where $m$, $P_b$, $e$ and $a$ are the total mass, orbital period, orbital eccentricity and semi-major axis, respectively. 
Using the Kepler's third law, the expression \eqref{PK_ome_dot} for the periastron advance is further simplified and summarized in Eqs.\eqref{PK_p}.

\subsection{Time Delay}
\subsubsection{Einstein Delay}
The combined effect of gravitational and kinetic time dilation is so-called Einstein delay.
In a circular orbit, the Einstein delay can be absorbed as a constant parameter, and it is meaningless.
In an elliptical orbit, the Einstein delay is always changing with time due to a variation in the pulsar velocity and a change of the distance between the pulsar and its companion.

The Einstein delay in an elliptical orbit can be computed by the proper time at the pulsar's point of emission, 
\begin{eqnarray}
d{\tau}_p = dt \Big(-g_{\mu\nu}\frac{dx_p^{\mu}}{dt}\frac{dx_p^{\nu}}{dt}\Big)^{\frac{1}{2}},
\end{eqnarray}
where the subscript $p$ represents the pulsar.
Using the PN expressions in \eqref{PN_metric_scalar}, integrating the above equation, and dropping the constant terms, the result to first order is given by
\begin{subequations}
\begin{eqnarray}
{\tau}_p={t}-\gamma\sin{E},
\end{eqnarray}
with
\begin{eqnarray}
\gamma=\frac{\mathscr{G}Gm_c}{a}\Big(1+\frac{m_c}{m}\Big)\frac{P_b}{2\pi}e,
\end{eqnarray}
\end{subequations}
where $E$ is the eccentric anomaly of the orbit and $m_c$ is the companion mass.
The parameter $\gamma$ is the amplitude of Einstein delay, using the Kepler's third law, and it is rewritten as 
\begin{eqnarray}
{\gamma}=e\frac{P_b}{2\pi}\left(\frac{2\pi{\mathscr{G}Gm}}{P_b}\right)^{\frac{2}{3}}\frac{m_c}{m}\left(1+\frac{m_c}{m}\right),
\end{eqnarray}
which is identical to that of GR in the limit of $\ep\rightarrow0$ \cite{Will:2014aa}.

\subsubsection{Shapiro Delay}\label{Shapiro}
The retardation of light signal caused by the reduced coordinate velocity of light in a gravitational field is so-called Shapiro delay \cite{Shapiro:1964aa}.
In binary pulsar systems, the Shapiro delay is usually parameterized by \cite{Wex:2014aa}
\begin{eqnarray}
\Delta t_{\rm S}
&=& 2r\ln\Big[1-e\cos{E}-s\sin{\omega}(\cos{E}-e) \nonumber\\
&& - s\cos{\omega}(1-e^2)^{\frac{1}{2}}\sin{E}\Big],
\end{eqnarray} 
where $r$ and $s$ are called the range and shape of the Shapiro delay, and $\omega$ is the longitude of periastron.

The light signal travels along a null geodesic $g_{\mu\nu}dx^{\mu}dx^{\nu}=0$, which remains unchanged under the conformal transformation. 
In other words, photons do not couple to the scalar field in $f(R)$ gravity, because the electromagnetic energy-momentum tensor has a vanishing trace.
Using the PN expressions \eqref{PN_metric_scalar}, the equation of null geodesic translates into the coordinate velocity of light,
\begin{eqnarray}
c_{\gamma}(\mathbf{r}) 
\equiv \frac{\sqrt{dx^i dx^j \delta_{ij}}}{dt} 
= 1-2\sum_a\frac{Gm_a}{| \mathbf{r}-\mathbf{r}_a |} + \mathcal{O}(v^4).~~~~~
\end{eqnarray}
The Shapiro delay can be obtained by the integral $\Delta t_{\rm S} = \int dz/c_{\gamma}(\mathbf{r})$.
Clearly, $c_{\gamma}(\mathbf{r})$ is exactly the same as that in GR, which indicates that the Shapiro delay parameters are also the same as those in GR. 
Therefore, the range of the Shapiro delay is given by $r=r_{\gr}=Gm_c$ \cite{Will:2014aa}.
The shape of the Shapiro delay is defined by $s\equiv\sin{i}=x_p/a_p$, where $a_p$ and $x_p$ are the semi-major axis and projected semi-major axis of the pulsar orbit, and $i$ is the orbital inclination angle.
Using the Kepler's third law, the shape of the Shapiro delay is rewritten as 
\begin{eqnarray}
s={x_p}\left(\frac{2\pi}{{P_b}}\right)^{\frac{2}{3}}\frac{m^\frac{2}{3}}{{(\mathscr{G}G)^\frac{1}{3}}m_c}.
\end{eqnarray}

\subsection{Orbital Decay}
In the previous sections, the periastron advance and time delay only describe the conservative sector of the theory. 
In this section we focus on the dissipative effects, calculate the loss rates of the orbital energy and angular momentum from the emission of GWs predicted by $f(R)$, and derive their contributions to the orbital decay.

\subsubsection{Energy and Angular Momentum Fluxes}
The orbital decay due to GWs damping is very important for testing gravity \cite{Hulse:1975aa, Taylor:1982aa, Taylor:1989aa}, and its theoretical derivation is also the basis of GWs waveform calculation \cite{Liu:2018aa, Zhang:2017ab}.

In the far zone, the tensor and scalar fields can be decomposed as the perturbations about the Minkowski background and the scalar background, i.e., $g_{\mu\nu}=\eta_{\mu\nu}+h_{\mu\nu}$ and $\phi=\ph+\varphi$.
Using these and imposing the transverse-traceless (TT) gauge on the Lagrangian density \eqref{Lagrangian_fR_E}, expanding to quadratic order in the perturbations $h_{\mu\nu}$ and $\varphi$, the Lagrangian densities of the tensor and scalar GWs are given by
\begin{eqnarray}
\mathcal{L}_{T} 
&=& -\frac{\Mpl^2}{8}\partial_{\mu} h_{ij}^{\rm TT}\partial^{\mu} h_{ij}^{\rm TT}, \\
\mathcal{L}_{S} 
&=& -\frac{(\partial\varphi)^2}{2}-\frac{1}{2}m^2_s\varphi^2,
\end{eqnarray}
where ${h}^{\rm TT}_{ij}$ is the TT part of ${h}_{ij}$, and $m^2_{s}=d^2 V_{\rm eff}/d\phi^2|_{\ph}$ is the scalar field mass.
Energy and angular momentum are the conserved charges associated to time translation invariance and spatial rotation invariance, respectively.
The energy and angular momentum fluxes of the tensor and scalar GWs are derived directly from the above Lagrangians by investigating the Noether charges and currents, given by
\begin{subequations}
\label{E_L_fluxes}
\begin{eqnarray}
\label{tensor_Eflux}
\dot{E}_{T} 
&=& \frac{r^2}{32\pi G} \!\int\! {d\Omega} {\big\langle} \dot{h}^{\rm TT}_{ij}\dot{h}^{\rm TT}_{ij} {\big\rangle}, \\
\label{scalar_Eflux}
\dot{E}_{S} 
&=& - r^2 \!\int\! {d\Omega} {\big\langle} \dot{\varphi}\partial_r{\varphi} {\big\rangle}, \\
\label{tensor_Lflux}
\dot{L}^i_{T} 
&=& \epsilon^{ijk}\frac{r^2}{32\pi G} \!\int\! {d\Omega} {\big\langle} 2h_{jl}^{\rm TT}\dot{h}_{kl}^{\rm TT}-\dot{h}_{lm}^{\rm TT}x^j\partial_kh_{lm}^{\rm TT} {\big\rangle}, \qquad~\\
\label{scalar_Lflux}
\dot{L}^i_{S} 
&=& -\epsilon^{ijk}r^2 \!\int\! {d\Omega} {\big\langle} \dot{\varphi} x^j\partial_k\varphi {\big\rangle}, 
\end{eqnarray}
\end{subequations}
where the overdots denote time derivatives, the angular brackets stand for an average over an orbital period, $\Omega$ is the solid angle, and $\epsilon^{ijk}$ is the Levi-Civita symbol.
Obviously, the energy and angular momentum fluxes of the tensor GWs are exactly the same as those in GR.
In Eq.\eqref{tensor_Lflux}, the angular momentum flux of the tensor GWs comes from the contributions of the spin and orbital angular momentum of the tensor graviton.
In Eq.\eqref{scalar_Lflux}, the angular momentum flux of the scalar GWs comes only from the orbital angular momentum of the scalar GWs, because the scalar field is spin-0.

\subsubsection{Wave Solutions}
In the far zone, expanding the field equations \eqref{tensor_Eq} and \eqref{scalar_Eq} to linear order in the perturbations $h_{\mu\nu}$ and $\varphi$, and imposing the Lorentz gauge $\partial^{\mu}(h_{\mu\nu}-\frac{1}{2}\eta_{\mu\nu}h)=0$, the wave equations for the perturbations are given by \cite{Zhang:2017aa, Zhang:2019ac}
\begin{eqnarray}
\label{tensor_wave_eq}
&&\square{h}_{\mu\nu}=-16\pi G\big(T_{\mu\nu}-\frac{1}{2}\eta_{\mu\nu}T\big),
\\
\label{scalar_wave_eq}
&&(\square-m^2_s)\varphi=-\partial_{\varphi} T,
\end{eqnarray}
where $\square\equiv\eta^{\mu\nu}\partial_{\mu}\partial_{\nu}$ is the d'Alembertian of the flat space-time, and $T = \eta^{\mu\nu}T_{\mu\nu}$. 
Note that here, $T_{\mu\nu}$ is the energy-momentum tensor of the matter, the energy-momentum tensors of the perturbations do not contribute to the wave equations in the linear regime.
From the definition of $T_{\mu\nu}\equiv-(2/\sqrt{-g})\delta S_m/\delta g^{\mu\nu}$, using the matter action of $S_m=-\sum_{a}\int m_a(\phi)d\tau_a$ (see Eq.~\eqref{matter_Lag}), yields
\begin{eqnarray}
\label{Tuv_matter}
T^{\mu\nu}\!=(-g)^{\!-\! \frac{1}{2}} \! \sum_{a}m_a(\phi) u^{\mu}_{a} u^{\nu}_{a} ({u^{0}_{a}})^{\!-\!1}\delta^3 \! (\mathbf{r} - \mathbf{r}_a\!),\quad
\end{eqnarray}
where $u^{\mu}_a$ is the unit four-velocity of the $a$-th body.

By substituting a plane wave $\varphi\sim e^{ik^\lambda x_\lambda}$ into $(\square-m_s^2)\varphi=0$, yields the dispersion relation $\omega^2={\mathbf k}^2+m_s^2$, where $\omega$ and $\mathbf k$ are the frequency (energy) and wave vector of the scalar GWs, and $k^\lambda=(\omega,{\mathbf k})$.
It is clear that the scalar mode in $f(R)$ can be excited only when its the energy is greater than its mass.
In general, $m_s\ll\omega$ for compact binaries, because $m_s\sim10^{-12}{\rm Hz}$ for a scalar fifth force range on galactic scales ($\sim10{\rm kpc}$) and $\omega\sim10^{-3} {\rm Hz}$ for a typical binary pulsar with a 1 hour orbital period.
Therefore, the scalar field mass is neglected in the calculations below.

By using the Green's function method, the formal solutions of the wave equations are
\begin{eqnarray}
&&{h}_{ij}^{\rm TT}(t,\mathbf{r})=4G \Lambda_{ij,kl}(\mathbf{n}) {\int} d^3\mathbf{r'} \frac{T_{kl}(t-|\mathbf{r-r'}|,\mathbf{r'})}{|\mathbf{r-r'}|},~~~~
\\
&&\varphi(t,\mathbf{r})=\frac{1}{4\pi} {\int} d^3\mathbf{r'} \frac{\partial_{\varphi} T(t-|\mathbf{r-r'}|,\mathbf{r'})}{|\mathbf{r-r'}|},
\end{eqnarray}
where we have used ${h}_{ij}^{\rm TT}=\Lambda_{ij,kl}{h}_{kl}$ and $\Lambda_{ij,kl}\delta_{kl}=0$, $\Lambda_{ij,kl}(\mathbf{n})$ is the Lambda tensor as defined in \cite{Maggiore:2007aa}, and $\mathbf{n}=\mathbf{r}/r$ is a unit vector in the direction of $\mathbf{r}$.
Here the spatial (source point $\mathbf{r'}$) integration region is over the near zone, the field point $\mathbf{r}$ is in the far zone, i.e., $|\mathbf{r}'|\ll|\mathbf{r}|$, such that $|\mathbf{r-r'}|=r-\mathbf{r' \cdot n} + \mathcal{O}(r'^2/r)$.
Using this, the wave solutions can be expanded in the sum of a series of multipole moments,
\begin{eqnarray}
\label{tensor_wave_solu}
&&{h}_{ij}^{\rm TT}\!(t,\mathbf{r}) \!=\! \frac{4G}{r} \! \Lambda_{i\!j\!,k\!l}(\mathbf{n}) \!\!\sum_{{\ell}=0}^\infty \!\! \frac{1}{{\ell}!}\partial_t^{\ell} \!\! \int \!\! d^3\!\mathbf{r'} (\mathbf{r' \!\!\cdot\! n})^{\ell} T_{kl}\!(t \!-\! r, \mathbf{r'}\!),~~~~~
\\
\label{scalar_wave_solu}
&&\varphi(t,\mathbf{r})=\frac{1}{4\pi r}\sum_{{\ell}=0}^\infty\frac{1}{{\ell}!} \partial_t^{\ell} \int d^3\mathbf{r'}(\mathbf{r' \cdot n})^{\ell} \partial_{\varphi}T\big(t\!-\!r,\,\mathbf{r'}\big),
\end{eqnarray}
where $\partial_t^{\ell} \equiv (\partial/\partial t)^{\ell}$.

\subsubsection{Orbital Decay}
According to the balance law, the decay rates of the orbital energy and angular momentum equal to minus the energy flux and angular momentum flux of GWs of the emission, respectively.
For binary pulsar systems, substituting the multipole moment expressions \eqref{tensor_wave_solu} and \eqref{scalar_wave_solu} of the wave solutions into the expressions \eqref{E_L_fluxes} of the energy and angular momentum fluxes, performing a series of calculations, up to the 2.5PN order, and the decay rates of the orbital energy and angular momentum are given and summarized in Appendix \ref{appendix_EL}.
Keeping only the leading order terms in Eqs.~\eqref{total_E_dot} and \eqref{total_L_dot}, the results reduce to
\begin{eqnarray}
\label{total_E_dot1}
\dot{E}&=&
- \frac{32 G^4\mu^2m^3}{5 a^5} F(e)
- \frac{G^3\mu^2m^2}{6a^4}\frac{(1+\frac{1}{2}e^2)}{(1-e^2)^{\frac{5}{2}}} \ep_d^2, 
\\
\label{total_L_dot1}
\dot{L}&=&
- \frac{32 G^{\frac{7}{2}} \mu^2 m^{\frac{5}{2}}}{5 a^{\frac{7}{2}}}\frac{(1+\frac{7}{8}e^2)}{(1-e^2)^2}
- \frac{G^{\frac{5}{2}} \mu^2 m^{\frac{3}{2}}}{6a^{\frac{5}{2}}}\frac{\ep_d^2}{(1-e^2)}, \qquad
\end{eqnarray}
where $F(e)$ is defined in Eq.\eqref{F_e}, $\ep_p$ and $\ep_c$ are the scalar charges of the pulsar and its companion, $\ep_d \equiv \ep_p-\ep_c$, and $\mu \equiv m_1m_2/m$.
The orbital energy $E$ and the orbital angular momentum $L$ are related to the orbital semi-major axis $a$ and eccentricity $e$ through,
\begin{eqnarray}
E = - \frac{\mathscr{G}Gm\mu}{2a}, ~~~~ 
L^2=\mathscr{G}Gm\mu^2a(1-e^2).
\end{eqnarray}
Derivatives with respect to time yields
\begin{eqnarray}
\dot{a} \!=\! \frac{2a^2}{\mathscr{G}Gm\mu}\dot{E},~~~~
\dot{e} \!=\! \frac{a(1 \!-\! e^2)}{\mathscr{G}Gm\mu e}\Big[\dot{E} \!-\! \frac{(\mathscr{G}Gm)^{\frac{1}{2}}}{a^{\frac{3}{2}}(1 \!-\! e^2)^{\frac{1}{2}}}\dot{L}\Big],~~~~~
\end{eqnarray}
where $\dot{E}<0$ is a negative contribution to $\dot{e}$, and $\dot{L}<0$ is a positive contribution to $\dot{e}$. 
Substituting Eqs. \eqref{total_E_dot1} and \eqref{total_L_dot1} into the above expressions, and using the Kepler's third law, the decay rates of the orbital parameters $P_b$, $e$ and $a$ are given by
\begin{eqnarray}
\label{Pb_dot1}
\dot{P}_b &=&
-\frac{192\pi}{5} \! \Big( \frac{2{\pi}Gm}{P_b} \Big)^{\!\frac{5}{3}} \! \frac{\mu}{m} F(e) 
- \frac{2\pi^2 G\mu (1 \!+\! \frac{1}{2}e^2) \ep^2_d}{P_b (1 \!-\! e^2)^{\frac{5}{2}}}, 
\\
\label{e_dot1}
\dot{e} &=& 
-\frac{608\pi}{15} \! \Big( \frac{2{\pi}Gm}{P_b} \Big)^{\!\frac{5}{3}} \! \frac{{\mu} e (1 \!+\! \frac{121}{304}e^2)}{m P_b (1 \!-\! e^2)^{\frac{5}{2}}}
-\frac{\pi^2 G\mu e \ep_d^2}{P_b^2 (1 \!-\! e^2)^{\frac{3}{2}}}, \qquad
\\
\dot{a} &=& \frac{1}{3 \pi} \Big( \frac{2{\pi}Gm}{P_b} \Big)^{\frac{1}{3}} {\dot{P}_b}.
\end{eqnarray}
Here, the first and second terms are the quadrupole and dipole radiation.
It can be seen that the orbital decay for an asymmetric binary system is dominated by the dipole radiation and is typically faster than in GR.
The above results will return to the GR case when $\ep_p=\ep_c=0$.
In fact, most extended theories of GR include extra helicity-0 or helicity-1 degrees of freedom, both of which can open up new channels of dipole gravitational radiation in asymmetric binary systems \cite{Maggiore:2007aa}. 
Therefore, testing dipole radiation can also probe whether gravity includes these degrees of freedom.
Note that, the above expressions are also applicable to most theories of gravity with dipole radiation, and the only difference is that the model-dependent coefficients in dipole radiation are different.

\begin{figure}[!htbp]
\centering
\includegraphics[width=1\columnwidth]{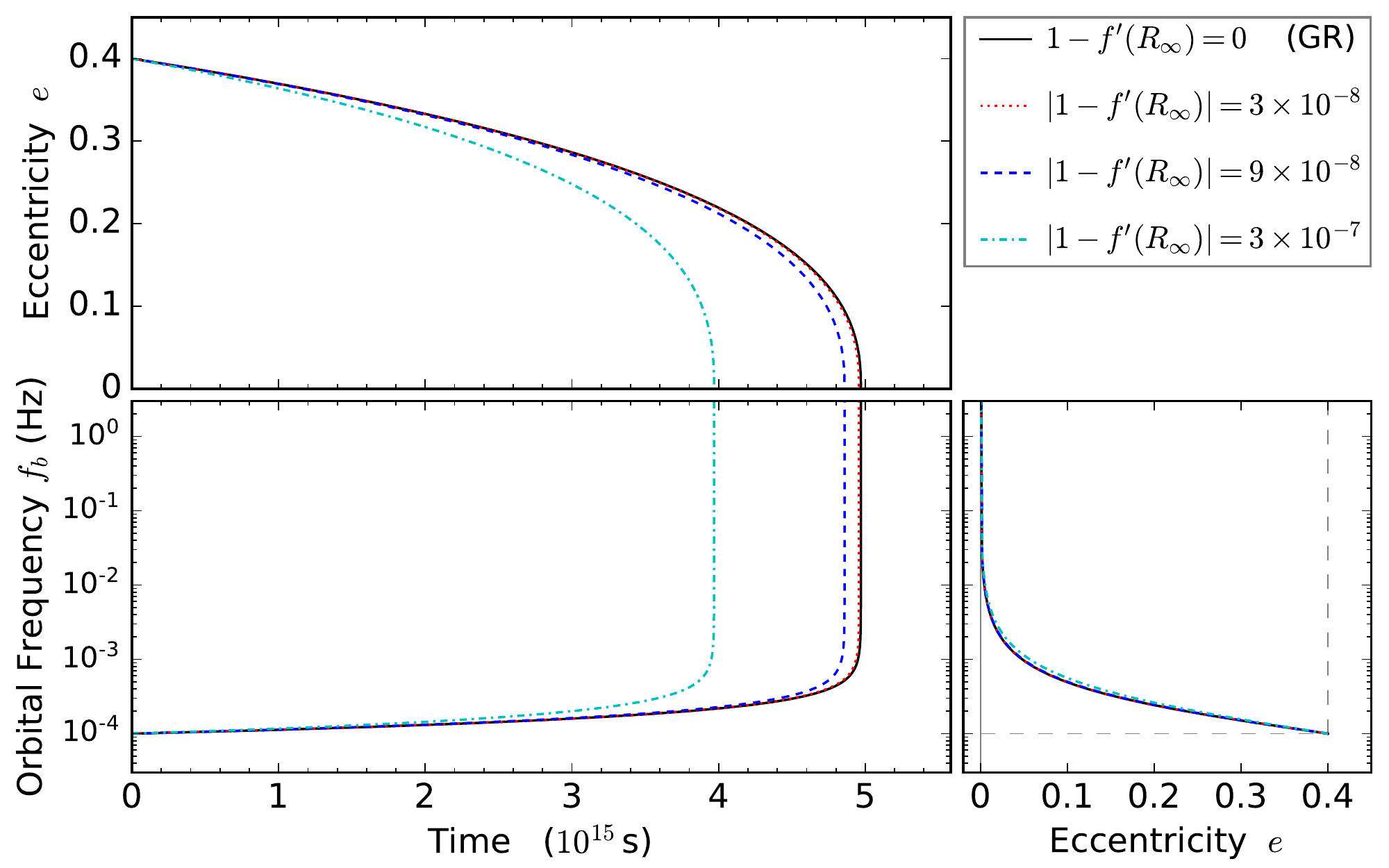}
\caption{
Temporal evolution of the orbital frequency and eccentricity of a $1.6M_\odot$ - $0.4M_\odot$ NS-WD binary system with compactnesses of $0.2$ - $5\times10^{-5}$ in $f(R)$ theories with different values of $\fR$.
The initial values of the orbital frequency and eccentricity are $10^{-4}$ Hz and 0.4.
}
\label{fig_fe}
\end{figure}

The orbital evolution can be obtained by solving the above system of nonlinear differential equations.
In Figure~\ref{fig_fe}, we show the temporal evolution of the orbital frequency ($f_b=1/P_b$) and eccentricity ($e$) of a NS-WD binary system in $f(R)$ theories with different values of $\fR$.
The temporal evolution is given by numerically solving Eqs.~\eqref{Pb_dot1} and~\eqref{e_dot1} for a $1.6M_\odot$ - $0.4M_\odot$ NS-WD binary system with compactnesses of $0.2$ - $5\times10^{-5}$ and an initial eccentricity of 0.4 and an initial orbital frequency of $10^{-4}$ Hz (correspond to an initial orbital period of 2.8 hours). 
Observe that the orbital frequency and eccentricity decay typically faster in $f(R)$ than in GR due to the presence of dipole radiation.

\subsection{Summary of PK parameters}
For convenience, the PK parameters are rewritten and summarized as follows:
\begin{subequations}
\label{PK_p}
\begin{eqnarray}
\label{PK_p1}
\dot{\omega}
&=& \dot{\omega}^{\gr}(1-\frac{1}{3} \ep_p\ep_c), \\
\label{PK_p2}
\gamma
&=& \gamma^{\gr}(1+\frac{1}{3}\ep_p\ep_c), \\
\label{PK_p3}
r
&=& r^{\gr}, \\
\label{PK_p4}
s
&=& s^{\gr}(1-\frac{1}{6}\ep_p\ep_c), \\
\label{PK_p5}
\dot{P_b}
&=& \dot{P}_b^{\gr}\Big[1+\frac{5}{192}\Big(\frac{P_b}{2{\pi}Gm}\Big)^{\frac{2}{3}}\frac{(1+\frac{1}{2}e^2) \ep_d^2}{(1-e^2)^{\frac{5}{2}}F(e)}\Big],\\
\label{PK_p6}
\dot{e}
&=& \dot{e}^{\gr}\Big[1+\frac{15}{1216}\Big(\frac{P_b}{2{\pi}Gm}\Big)^{\frac{2}{3}}\frac{(1-e^2) \ep_d^2}{1+\frac{121}{304}e^2}\Big], \\
\label{PK_p7}
\dot{a}
&=& \dot{a}^{\gr}\Big[1+\frac{5}{192}\Big(\frac{P_b}{2{\pi}Gm}\Big)^{\frac{2}{3}}\frac{(1+\frac{1}{2}e^2) \ep_d^2}{(1-e^2)^{\frac{5}{2}}F(e)}\Big], \qquad
\end{eqnarray}
with
\begin{eqnarray}
\label{PK_p8}
\ep_p\ep_c\!=\!\frac{3[\ln \! \fR]^2}{2\Phi_p\Phi_c},  \quad
\ep_d^2\!=\!3[\ln \! \fR]^2\frac{(\Phi_p\!-\!\Phi_c)^2}{2\Phi_p^2\Phi_c^2}, \qquad~
\end{eqnarray}
\end{subequations}
where the superscript GR denotes the GR values of the PK parameters (see Appendix \ref{appendix_GRPK}).
The periastron advance, Einstein delay and Shapiro delay are the PK effects of 1PN, 1PN and 1.5PN, respectively. 
The orbital decay rates come from the contributions of 2.5PN quadrupole radiation and 1.5PN dipole radiation.
In binary pulsars, through timing analysis, in general, the measurement of $\dot{P}_b$ is more accurate than that of $\dot{e}$ (or $\dot{a}$), and hence the constraint on the theory from $\dot{P}_b$ is generally more stringent.
Therefore, in binary pulsars, $\dot{P}_b$ and the first four PK parameters in Eqs.\eqref{PK_p} are usually used to test GR.

\section{Binary Pulsar Tests}\label{section4}
In this section, we study how to place constraints on $f(R)$ gravity with binary pulsar observations.

\subsection{Binary Pulsars}
\begin{table*}[!htbp]
\centering
\caption{Timing model parameters for three binary pulsar systems. Numbers in parentheses represent 1$\sigma$ (68.3\%) uncertainties in the last quoted digit. $^{\rm a}$The masses are derived by assuming that GR is valid. $^{\rm b}$The WD radius is derived by the WD mass-radius relation \cite{Hamada:1961aa}. $^{\rm c}$The NS radius is derived by assuming APR EoS is valid \cite{Akmal:1998aa}.}
\label{tab_PSRs}
\renewcommand{\arraystretch}{1.1}
\begin{tabular}{lrrrr}
\hline\hline
PSR Name & J1141$-$6545 \cite{Bhat:2008aa,Ord:2002aa}
& J1738$+$0333 \cite{Freire:2012aa}
& J0348$+$0432 \cite{Antoniadis:2013aa}
\\ \hline
Orbital period, $P_b$ (days)&0.1976509593(1)&0.3547907398724(13)&0.102424062722(7)\\
Projected semi-major  axis, $x_p$ (s)&1.858922(6)&0.343429130(17)&0.14097938(7)\\
Eccentricity, $e$&0.171884(2)&$0.34(11)\times10^{-6}$&$0.24(10)\times10^{-5}$\\
Periastron advance, $\dot\omega$ (deg/yr)&5.3096(4)& ... & ... \\
Einstein delay, $\gamma$ (ms)&0.773(11)& ... & ... \\
Observed $\dot P_b$, $\dot P_b^{\text{obs}}$ $(10^{-13})$&$-4.03(25)$&$-0.170(31)$&$-2.73(45)$\\
Intrinsic $\dot P_b$, $\dot P_b^{\text{int}}$ $(10^{-13})$&$-4.01(25)$&$-0.259(32)$&$-2.73(45)$\\
Shapiro delay, $s$ &0.97(1)& ... & ... \\
Mass ratio, $q=m_{\N}/m_{\W}$& ... &8.1(2)&11.70(13)\\
WD mass, $m_{\W}$ ($M_{\sun}$)&$1.02(1)^{\rm a}$&${0.181^{+0.008}_{-0.007}}$&$0.172(3)$ \\
NS mass, $m_{\N}$ ($M_{\sun}$)&$1.27(1)^{\rm a}$&${1.46^{+0.06}_{-0.05}}^{\rm a}$&$2.01(4)^{\rm a}$ \\
WD radius, $R_{\W}$ ($R_{\sun}$) &   $0.0080(1)^{\rm b}$   & $0.037_{-0.003}^{+0.004}$ & $0.065(5)$  \\
NS radius, $R_{\N}$ (km) &   $11.391(2)^{\rm c}$   & $11.35(2)^{\rm c}$ & $10.89(8)^{\rm c}$  \\
\hline\hline
\end{tabular}
\end{table*}

Binary pulsars are crucial as the first indirect detectors of GWs \cite{Hulse:1975aa,Taylor:1982aa,Taylor:1989aa}.
Binary pulsars possess extreme gravitational environment, making them very useful tools for testing strong-field gravity.
The orbital decay in $f(R)$ gravity is dominated by the dipole radiation, which depends on the difference in the compactnesses of pulsar and its companion (see Eqs.~\eqref{PK_p}).
Therefore the asymmetric systems like NS-WD binaries are one of the ideal targets to test $f(R)$ gravity.
Moreover, although the theory parameter $\fR$ is degenerate with the compactnesses of binaries, the degeneracy can be broken by the radii of binaries.
Therefore, in the PK parameters, there are only three independent parameters $m_p$, $m_c$ and $\fR$ to be determined. 
For all these reasons, testing $f(R)$ gravity by binary pulsars requires that they can provide at least three PK observables, including the intrinsic $\dot{P}_b^{\rm int}$~\footnote{The intrinsic $\dot{P}_b^{\rm int}$ can be obtained from the observed value of $\dot{P}_b^{\rm obs}$ by subtracting two main effects: the differential galactic acceleration \cite{Damour:1991aa} and the Shklovskii effect \cite{Shklovskii:1970aa}.
} caused by gravitational radiation damping.
Based on the above analysis, we consider the following three NS-WD systems: PSRs J1141$-$6545 \cite{Bhat:2008aa,Ord:2002aa}, J1738$+$0333 \cite{Freire:2012aa} and J0348$+$0432 \cite{Antoniadis:2013aa}. 
Among these three NS-WD systems only the latter two systems provide the measured value of WD radius.
The radius of the WD in the first system is estimated by using the WD mass-radius relation \cite{Hamada:1961aa}.
For each NS in these three systems, the NS radius is estimated by using the mass-radius relation derived from the equations of state (EoS) based on the Akmal, Pandharipande, and Ravenhall (APR)~\footnote{So far, the NS EoS is not fully known. Here we consider the APR model and assume that it is valid.} model \cite{Akmal:1998aa}.
The relevant parameters for these three systems are listed in Table \ref{tab_PSRs}.

\subsection{Method and Results}
\begin{figure}[!htbp]
\centering
\includegraphics[width=1\columnwidth]{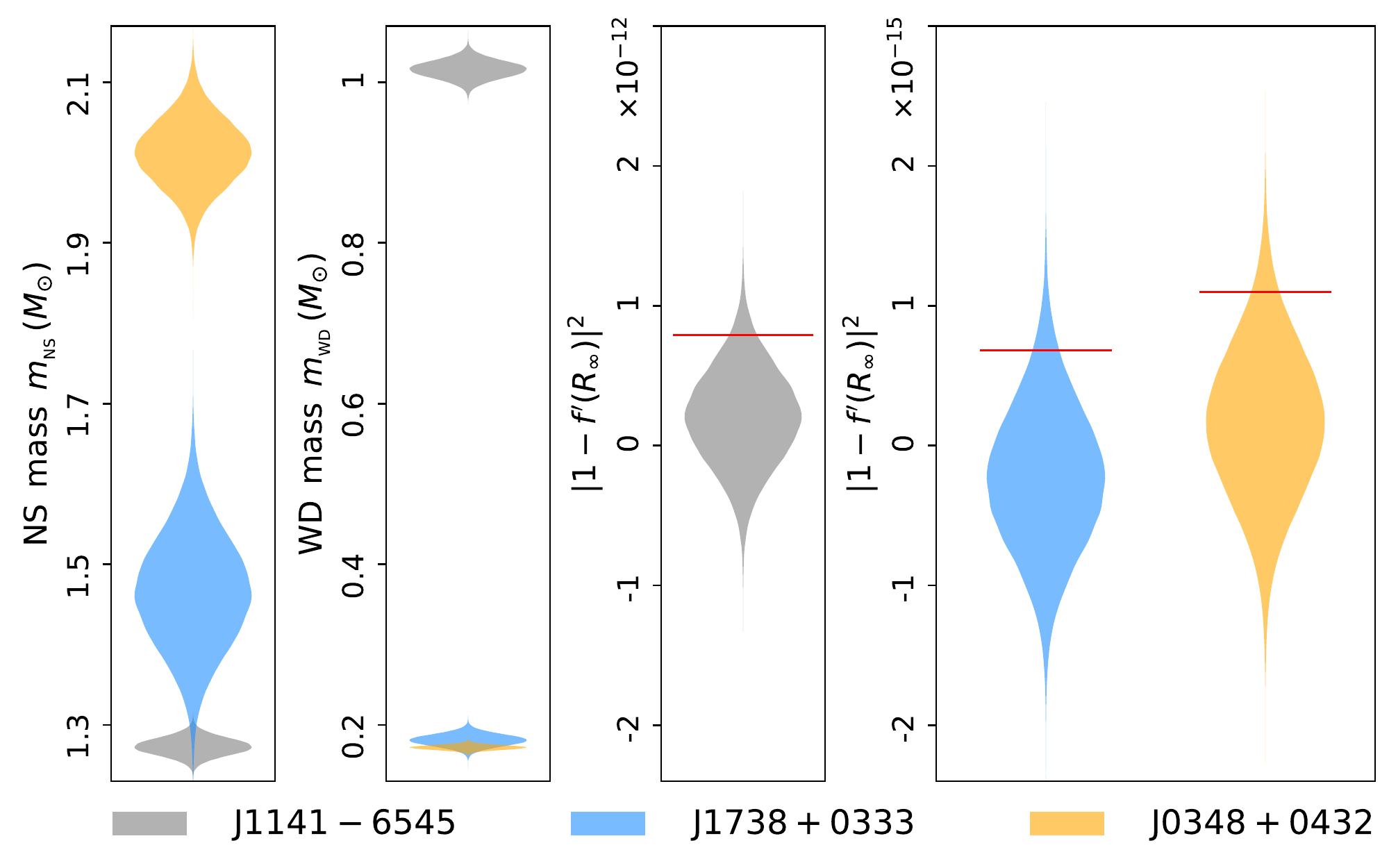}
\caption{Violin plots of parameters $m_{\N}$, $m_{\W}$ and $|1-\fR|^2$ for binary pulsars. 
PSRs J1141$-$6545, J1738$+$0333 and J0348$+$0432 are represented in gray, blue and orange, respectively.
The red lines represent the 95.4\%~CL upper bound.}
\label{fig_m1m2fR_Violin}
\end{figure}

We perform a Monte-Carlo simulation to determine these three unknown parameters $m_p$, $m_c$ and $\fR$ for each of these systems mentioned above.
In this simulation, the input quantities are mainly the PK observables, and each of them is randomly sampled from a normal distribution with mean and standard deviation equal to its observed value and 1-$\sigma$ uncertainty.
Then, these unknown parameters as the output quantities are estimated by numerically solving the system of equations \eqref{PK_p} of the PK parameters. 
This process is repeated $10^6$ times to construct the histograms of these unknown parameters and determine their median values and uncertainties.
The results are shown in Fig.~\ref{fig_m1m2fR_Violin} and Table~\ref{tab_m1m2fR}.

%
%
\begin{table}[!htbp]
\centering
\caption{Parameters $m_{\N}$, $m_{\W}$, and upper bound on $|1-\fR|$ at 95.4\%~CL for binary pulsars.}
\label{tab_m1m2fR}
\renewcommand{\arraystretch}{1.1}
\begin{tabular}{p{2cm}<{\raggedright}p{1.5cm}<{\raggedleft}p{2cm}<{\raggedleft}p{2.5cm}<{\raggedleft}}
\hline\hline
PSR Name        &   $m_{\N}\,(M_{\sun})$        &  $m_{\W}\,(M_{\sun})$         &   $|1-\fR|\le$      \\
\hline
J1141$-$6545    &   $1.27(1)$         &  $1.02(1)$          &  $8.9\times10^{-7}$            \\
J1738$+$0333   &   $1.47(7)$         &  $0.181(8)$        &  $2.6\times10^{-8}$           \\
J0348$+$0432  &   $2.01(4)$         &  $0.172(3)$        &  $3.3\times10^{-8}$           \\
\hline\hline
\end{tabular}
\end{table}
%

\begin{figure*}[!htbp]
\centering
\subfigure[PSR J1141$-$6545]{
\label{fig_a} 
\includegraphics[width=5.8cm, height=5.8cm]{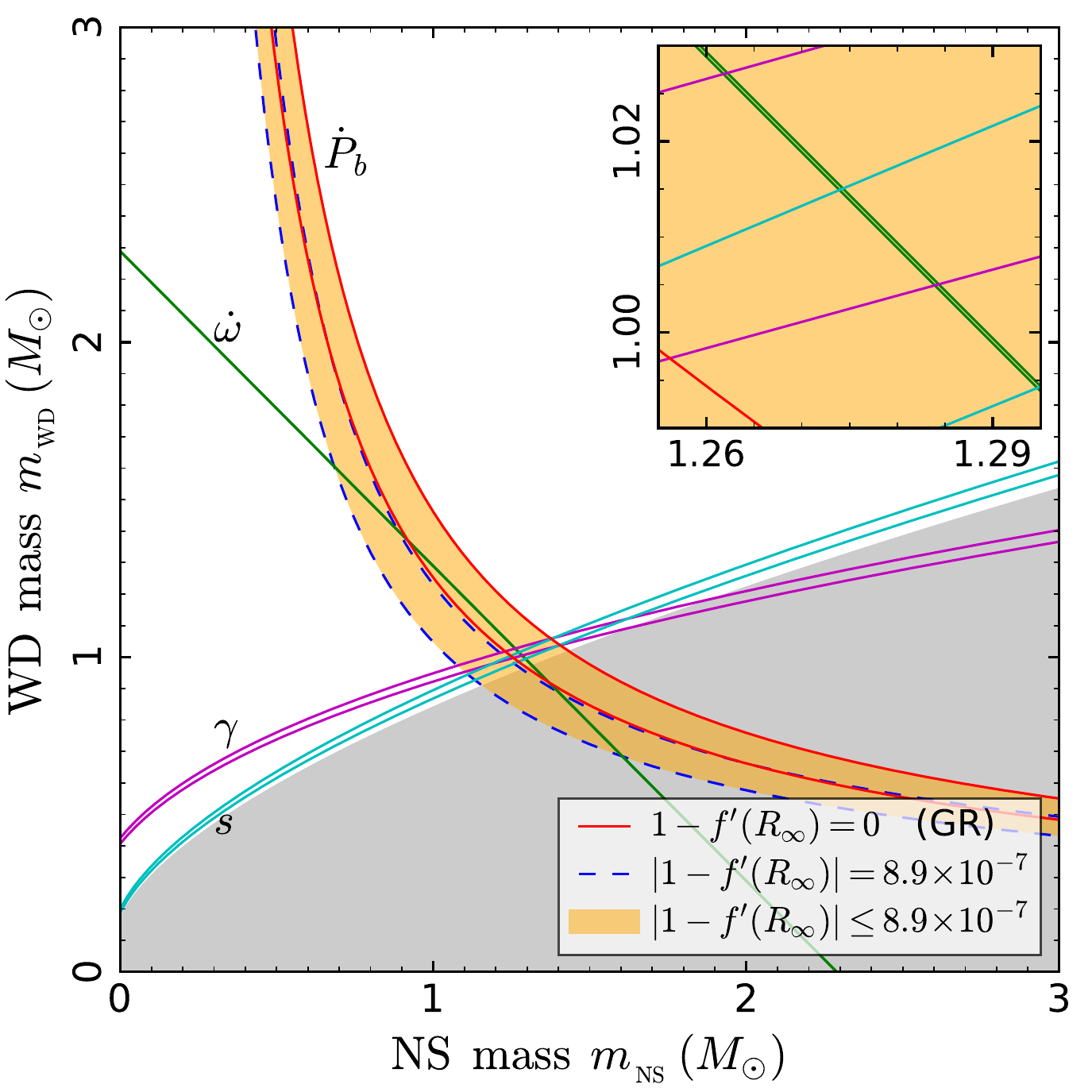}}
\subfigure[PSR J1738$+$0333]{
\label{fig_b} 
\includegraphics[width=5.8cm, height=5.8cm]{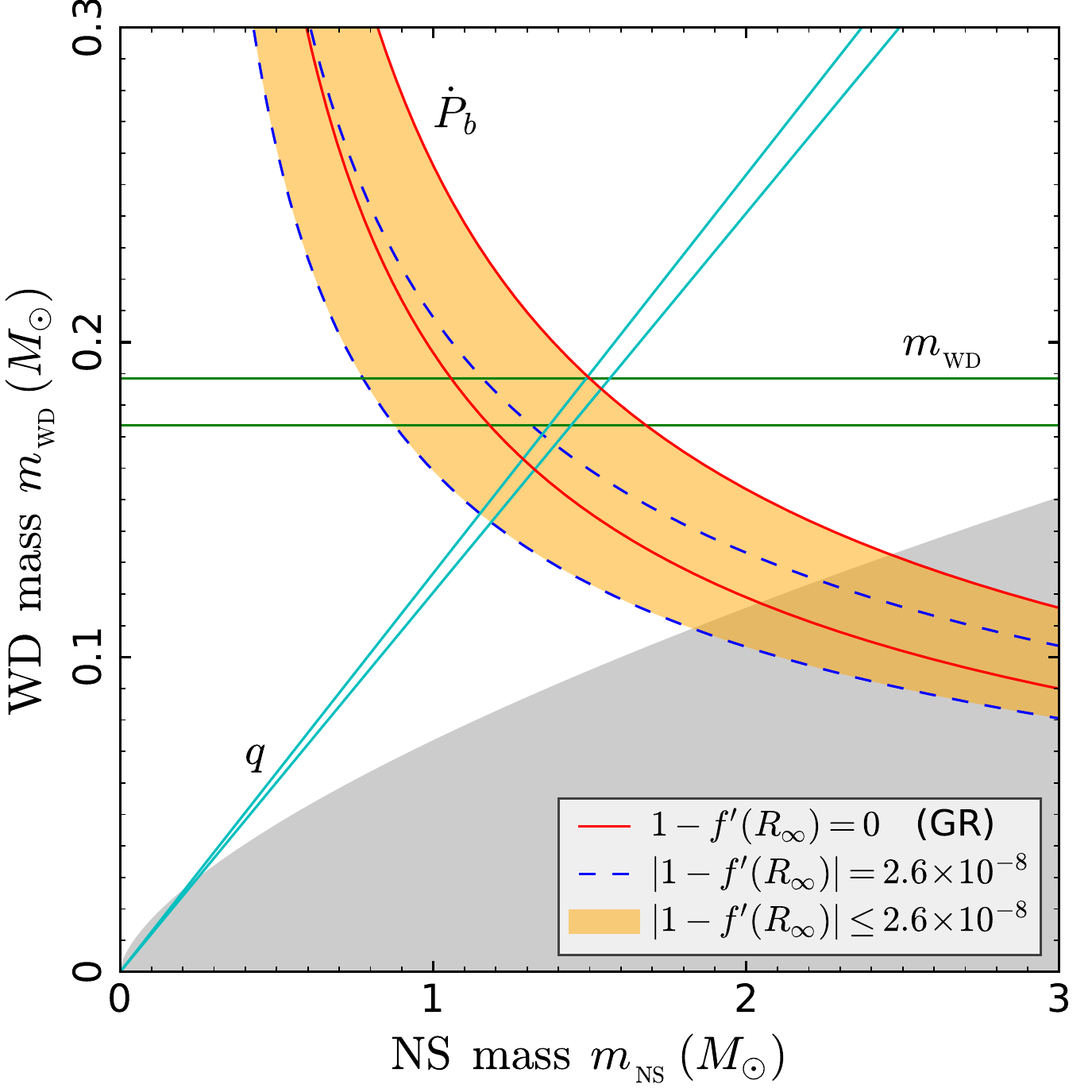}}
\subfigure[PSR J0348$+$0432]{
\label{fig_c} 
\includegraphics[width=5.8cm, height=5.8cm]{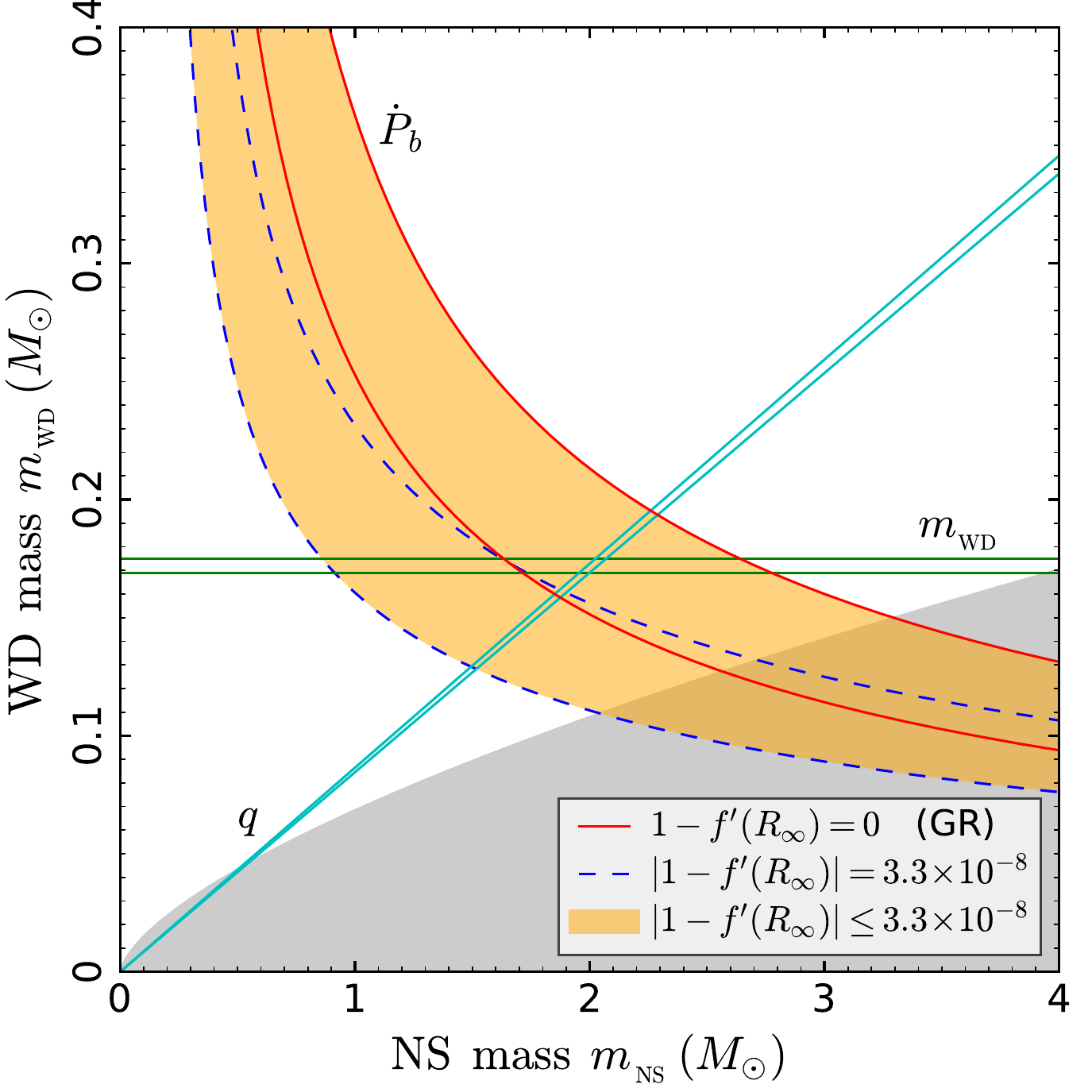}}
\caption{Mass-mass diagrams for the NS-WD PSRs J1141$-$6545, J1738$+$0333 and J0348$+$0432.
In (a), for $\dot\omega$, $\gamma$ and $s$, the dashed curves ($f(R)$) are covered by the solid curves (GR).
In (b) and (c), $q$ and $m_{\W}$ are independent of specific gravity theories.
The width of each curve represents $\pm1\sigma$ error bounds.
The gray regions are ruled out by the condition $s\equiv\sin{i}\le1$.}
\label{fig_MM}
\end{figure*}

PSR J1141$-$6545 is a 394 ms pulsar in a 4.74 hours elliptical orbit with a WD companion.
This system provides the four PK observables $\dot{P}_b^{\rm int}$, $\dot\omega$, $\gamma$ and $s$.
The first three observables are used to compute the three unknown parameters $m_{\N}$, $m_{\W}$ and $\fR$, and the last is the only one test on $f(R)$.
By performing a Monte-Carlo simulation, the unknown parameters for this system are derived and shown in Fig.~\ref{fig_m1m2fR_Violin} and Table~\ref{tab_m1m2fR}.
This system gives the upper bound on $|1-\fR|$ of $8.9\times10^{-7}$ at 95.4\%~confidence level (CL).
These results imply the Shapiro delay shape $s=0.96(1)$ (68.3\%~CL) in $f(R)$, which agrees with its observed value 0.97(1) (see Table \ref{tab_PSRs}).

PSRs J1738$+$0333 and J0348$+$0432 are millisecond pulsars in low-eccentricity orbits with low-mass WD companions.
Each of these two systems provides only the three observables $\dot{P}_b^{\rm int}$, WD mass $m_{\W}$ and mass ratio $q$.
Using these observables and performing Monte-Carlo simulations, we obtain the upper bounds on $|1-\fR|$ of $2.6\times10^{-8}$ and $3.3\times10^{-8}$ at 95.4\%~CL from PSRs J1738$+$0333 and J0348$+$0432 (see Fig.~\ref{fig_m1m2fR_Violin} and Table~\ref{tab_m1m2fR}), respectively.
These results rule out significant strong-field deviations of gravity from GR, and confirm that GR is a correct theory of gravity for asymmetric systems of strong gravity.
PSR J1738$+$0333 is the most constraining binary pulsar for testing $f(R)$ in these binary pulsar systems.

The mass-mass diagrams for PSRs J1141$-$6545, J1738$+$0333 and J0348$+$0432 are shown in Figs.~\ref{fig_a}, \ref{fig_b} and \ref{fig_c}, respectively.
These constraints on $m_{\N}$ and $m_{\W}$ in GR (solid) and in $f(R)$ (dashed) are based on the observables of the PK parameters, WD mass and mass ratio.
The PK constraint curves in $f(R)$ (dashed) are obtained by giving the deviation parameter $|1-\fR|$ an upper limit (see Table~\ref{tab_m1m2fR}). 
In Fig.~\ref{fig_a}, for a very small value of $8.9\times10^{-7}$ of $|1-\fR|$, the $\dot\omega$, $\gamma$ and $s$ constraint curves in $f(R)$ (dashed) are exactly covered by those in GR (solid).
The $\dot P_b$ constraint curves in $f(R)$ (blue dashed) are significantly different from that in GR (red solid), because the stronger dipole radiation appears in $f(R)$.
Therefore, the constraint on $f(R)$ from PSR J1141$-$6545 mainly comes from the measured value of $\dot{P}_b^{\rm int}$.
In Figs.~\ref{fig_b} and \ref{fig_c}, the WD mass $m_{\W}$ and mass ratio $q$ are theory-independent, therefore the constraints on $f(R)$ from PSRs J1738$+$0333 and J0348$+$0432 only come from the measured value of $\dot{P}_b^{\rm int}$.
These constraints from $\dot{P}_b^{\rm int}$ exclude significant dipole radiation deviations, which indicates no solid evidence of the existence of helicity-0 or helicity-1 degrees of freedom.

\section{Conclusions}\label{section5}
Chameleon $f(R)$ gravity is a natural alternative to GR.
In this paper, in the framework of chameleon $f(R)$, we studied the full PK effects of binary pulsars, and constrained the theory by using the observed PK parameters of NS-WD binary pulsar systems.
The PK effects in chameleon $f(R)$ differ from those in GR and the deviations are quantified by a combination of theory parameter $\fR$ and star's compactness.
Because of the degeneracy between them, the theory parameter cannot be constrained alone. 
The parameter degeneracy is broken by using the mass-radius relation, which allows us to place constraints on theory parameter.
The temporal evolution of the orbital period and eccentricity is typically faster than in GR due to the emission of dipole radiation in chameleon $f(R)$.
We used the three NS-WD binary pulsars to place constraints on chameleon $f(R)$ by performing Monte-Carlo simulations.
These constraints can also be thought of as tests of dipole radiation, which can probe whether GWs include extra helicity-0 or helicity-1 polarization states.
The results show that PSR J1738$+$0333 is the most constraining binary pulsar for testing chameleon $f(R)$ in these systems.
The significant strong-field deviations from GR are excluded by binary pulsar tests.
All tests show good agreement with GR, which indicates that GR is correct for asymmetric systems of strong gravity.

\acknowledgments{
We appreciate the helpful discussion with Dan Wang. 
This work is supported by the National Natural Science Foundation of China (NSFC) Grant No.11903033 and the Fundamental Research Funds for the Central Universities under Grant No.WK2030000036.
}

\appendix

\section{Chameleon Constraints on $f(R)$}\label{appendix_cfR}
In chameleon $f(R)$ gravity, the chameleon mechanism allows the theory to escape the tight solar system tests.
Using Eq.\eqref{conf_fR_ST}, the constraint conditions \eqref{fR_V_const} of the chameleon mechanism can be translated into the following constraints on the functional form of $f(R)$ \cite{Brax:2008aa}.
\begin{eqnarray}
\frac{d V}{d\phi} &=& \frac{\Mpl}{\sqrt{6} f'^2}[Rf'-2f]<0, \nonumber\\
\frac{d^2 V}{d\phi^2} &=& \frac{1}{3}\Big[\frac{R}{f'}+\frac{1}{f''}-\frac{4f}{f'^2}\Big]>0,\\
\frac{d^3 V}{d\phi^3} &=& \frac{2}{3 \sqrt{6} \Mpl}\Big[\frac{3}{f''}+\frac{f'f'''}{f''^3}+\frac{R}{f'}-\frac{8f}{f'^2}\Big]<0, \nonumber
\end{eqnarray}
where a prime denotes differentiation with respect to $R$, and $f\equiv f(R) $.

\section{PN Solutions}\label{appendix_PN}
Here we derive the PN solutions of the field equations in the near zone.
In the PN formalism \cite{Will:1993aa,Will:2014aa}, the tensor and scalar fields are decomposed as 
\begin{align}
\begin{split}
g_{00} &= -1+\accentset{(2)}h_{00}+\accentset{(4)}h_{00}+...,\\
g_{0j} &= \accentset{(3)}h_{0j}+...,\\
g_{ij} &= \delta_{ij}+\accentset{(2)}h_{ij}+...,\\
\phi &= \ph+\accentset{(2)}{\varphi}+\accentset{(4)}{\varphi}+...,
\end{split}
\end{align}
where the superscript $(n)$ means that the quantity is of order $\mathcal{O}(v^n)$, and $\ph$ is the physical vacuum of the scalar field in the background (i.e., the scalar background) which depends on the background density.

By using the matter Lagrangian \eqref{matter_Lag}, performing the PN expansions of the field equations \eqref{tensor_Eq} and \eqref{scalar_Eq}, and imposing the PN gauge ($h^{\mu}_{i,\mu}-\frac{1}{2}h^{\mu}_{\mu,i}=0\label{PN_gauge_i}$ and $h^{\mu}_{0,\mu}-\frac{1}{2}h^{\mu}_{\mu,0}=-\frac{1}{2}h_{00,0}\label{PN_gauge_0}$) \cite{Will:1993aa,Will:2014aa}, the PN field equations are given by
\begin{subequations}
\begin{eqnarray}
&& \nabla^2\ac{(2)}h_{00}=-8\pi G\sum_am_a\delta^3(\mathbf{r}-\mathbf{r}_a),\\
&& \nabla^2\ac{(2)}h_{ij}=-8\pi G\delta_{ij}\sum_am_a\delta^3(\mathbf{r}-\mathbf{r}_a),\\
&& \nabla^2\ac{(3)}h_{0j}+\frac{1}{2}\ac{(2)}h_{00,0j}=16\pi G\sum_am_av_a^j\delta^3(\mathbf{r}-\mathbf{r}_a),\\
&& \nabla^2\ac{(4)}h_{00}+\frac{1}{2}\nabla^2\ac{(2)}h_{00}^2-\ac{(2)}h_{00}\nabla^2\ac{(2)}h_{00}-\ac{(2)}h_{jk}\ac{(2)}h_{00,jk}=\\
&& - 8\pi G \sum_a m_a\delta^3(\mathbf{r} - \mathbf{r}_a) \Big( \frac{3}{2}v^2_a - \accentset{(2)}h_{00} - \frac{1}{2}\accentset{(2)}h_{ij}\delta_{ij} + s_a\frac{\accentset{(2)}\varphi}{\ph}\Big),\nonumber\\
&& \square(\accentset{(2)}\varphi+\accentset{(4)}\varphi)=
\frac{8\pi\Mpl^2}{\ph}\sum_aGm_as_a\delta^3(\mathbf{r}-\mathbf{r}_a) \\
&& \times \bigg[1-\frac{1}{2}v_a^2-\sum_{b\ne a}\frac{Gm_b}{r_{b}}-2\frac{s'_a}{s_a}\Big(\frac{\Mpl}{\ph}\Big)^2\sum_{b\ne a}\frac{Gm_bs_b}{r_{b}}\bigg],\nonumber
\end{eqnarray}
\end{subequations}
where $v_a$ is the velocity of the $a$-th body, and $r_a=\left|\mathbf{r}-\mathbf{r}_a(t)\right|$. 
The mass $m_a\equiv m_a(\ph)$ is the inertial mass at $\ph$, and
\begin{eqnarray}
\label{sensitivities_s}
s_a\equiv\frac{\partial(\ln m_a)}{\partial(\ln \phi)}\bigg|_{\ph},~~
s'_a\equiv s_a^2 \!-\! s_a \!+\! \frac{\partial^2(\ln m_a)}{\partial(\ln \phi)^2}\bigg|_{\ph},~~~
\end{eqnarray}
are respectively the first and second sensitivities \cite{Eardley:1975aa, Alsing:2012aa}, which characterize how the gravitational binding energy of a strongly self-gravitating body responds to its motion relative to the extra fields. 
Note that here we have neglected the scalar field mass $m_s$ of cosmological scales and the potential $V(\phi)$ corresponding to the dark energy, since these effects are very weak in the near zone.

Solving the above system of equations, and summing the relevant components, the PN solutions of the field equations are
\begin{eqnarray}\label{PN_metric_scalar}
g_{00} 
&=& -1+2\sum_a\frac{Gm_a}{r_a}-2\bigg(\sum_{a}\frac{Gm_a}{r_a}\bigg)^2+3\sum_a\frac{Gm_av_a^2}{r_a}\nonumber\\
&& -2\sum_a\sum_{b\ne a}\frac{G^2m_am_b}{r_ar_{ab}}\left(1+\frac{1}{2}\ep_a\ep_b\right)+\mathcal{O}(v^6), \nonumber\\
g_{0j}
&=& - \frac{7}{2} \sum_a \frac{Gm_av_a^j}{r_a} - \frac{1}{2} \sum_{a} \frac{Gm_a}{r_a^3}(\mathbf{r}_a \cdot  \mathbf{v}_a) (r^j - r_a^j) \nonumber\\
&& + \mathcal{O}(v^5), \\
g_{ij}
&=& \delta_{ij} \left(1+2\sum_a\frac{Gm_a}{r_a}\right)+\mathcal{O}(v^4), \nonumber\\
\varphi
&=& -\Mpl\sum_a\frac{Gm_a\ep_a}{r_a}\bigg[1-\frac12v_a^2-\sum_{b\ne a}\frac{Gm_b}{r_{ab}} \nonumber\\
&& -\frac{s'_a}{s_a}\frac{\Mpl}{\ph}\sum_{b\ne a}\frac{Gm_b\ep_b}{r_{ab}}+\frac{r_a}{2}\frac{\partial^2r_a}{\partial t^2}\bigg]+\mathcal{O}(v^6), \nonumber
\end{eqnarray}
where $r_{ab}=\left|\mathbf{r}_a(t)-\mathbf{r}_b(t)\right|$. 
Here, the quantity $\ep_a$ is usually called the scalar charge, and connects with the sensitivity $s_a$ by $\ep_a={2\Mpl}s_a/{\ph}$ \cite{Zhang:2017aa}.
For a static spherically symmetric source of homogeneous density, the scalar charge is given by  \cite{Zhang:2016aa}
\begin{eqnarray}
\label{epsilon_a}
\ep_a = \frac{\ph-\phi_a}{\Mpl\Phi_{a}},
\end{eqnarray}
where $\phi_a$ is the position of the effective potential minimum inside the $a$-th body, and $\Phi_a=Gm_a/R_a$ is the compactness of the $a$-th body and $R_a$ is its radius.
Note that $\phi_a$ is generally inversely correlated to the matter density \cite{Zhang:2016aa}. 
For a compact object, its density is always much larger than the background density, and therefore there are $\ph\gg\phi_{a}$ and $\ep_a\simeq\ph/(\Mpl\Phi_a)$.

\section{Orbital Energy and Angular Momentum Decays}\label{appendix_EL}
In $f(R)$ gravity, the decay rates of the orbital energy and angular momentum are summarized as follows:
\begin{eqnarray}
\label{total_E_dot}
\dot{E}
&=& -\frac{G^3\mu^2m^2(1+\frac{1}{2}e^2)}{6a^4(1-e^2)^{\frac{5}{2}}}\ep^2_d
-\frac{G^4\mu^2m^3}{a^5(1-e^2)^{\frac{7}{2}}} \times\!\bigg\{
\nonumber\\
&& \frac{32}{5}\!\Big(1\!+\!\frac{73}{24}e^2\!+\!\frac{37}{96}e^4\Big)\!\Big(1\!+\!\frac{3}{2}\ep_1\ep_2\Big) \!+\! \frac{e^2}{4}\!\big(1\!+\!\frac{e^2}{4}\big)\ep_m^2
\nonumber\\
&& -\ep_d\!\Big[\!\frac{\ep_{d1}\!+\!\ep_{d2}}{3}\!+\!\big(\ep_{d1}\!+\!\frac{13}{6}\ep_{d2}\big)e^2\!+\!\big(\frac{\ep_{d1}}{8}\!+\!\frac{5}{12}\ep_{d2}\big)e^4\Big]
\nonumber\\
&& +\frac{8}{15}\big(1+\frac{99}{32}e^2+\frac{51}{128}e^4\big)\ep^2_q-\frac{e^2}{6}\big(1+\frac{e^2}{4}\big)\ep_m\ep_q
\nonumber\\
&& -\frac{1}{30}\big(1-18e^2-\frac{39}{8}e^4\big)\ep_d\ep_o\bigg\}, \\
\label{total_L_dot}
\dot{L}
&=& -\frac{G\mu^2(Gm)^{\frac{3}{2}}}{6a^{\frac{5}{2}}(1\!-\!e^2)}\epsilon_d^2 \!-\!\frac{G\mu^2(Gm)^{\frac{5}{2}}}{a^{\frac{7}{2}}(1\!-\!e^2)^2} \times \!\bigg\{\!\frac{32}{5}\!\big(1\!+\!\frac{7}{8}e^2\big)
\nonumber\\
&& \times \big(1\!+\!\frac{5}{4}\ep_1\ep_2\big) \!-\!\frac{\epsilon_d\epsilon_o}{60}\!\big(2\!-\!17e^2\big) \!+\! \frac{\epsilon_q^2}{15}\!\big(8\!+\!7e^2\big)
\nonumber\\
&& -\frac{1}{6}\Big[2\big(\epsilon_{d1}+\epsilon_{d2}\big)+e^2\big(\epsilon_{d1}+4\epsilon_{d2}\big)\Big]\epsilon_d
\bigg\},
\end{eqnarray}
where we have defined
\begin{eqnarray}
\ep_m 
&\equiv& \ep_1+\ep_2+\frac{\ep_1m_2+\ep_2m_1}{m}, \quad
\ep_d \equiv \ep_2-\ep_1, \qquad \nonumber\\
\ep_{d1} 
&\equiv& \frac{\ep_2m_1-\ep_1m_2}{m}, \qquad
\ep_{d2} \equiv \frac{\ep_2m_1^2-\ep_1m_2^2}{2m^2}, \\
\ep_q 
&\equiv& \frac{\ep_2m_1+\ep_1m_2}{m},  \qquad~
\ep_{o}\equiv\frac{\ep_2m_1^2-\ep_1m_2^2}{m^2}. \nonumber
\end{eqnarray}
Here the subscripts $m$, $d$, $q$ and $o$ denote monopole, dipole, quadrupole and octupole, respectively. 
In Eqs.~\eqref{total_E_dot} and \eqref{total_L_dot}, the first term is the scalar dipole radiation of 1.5PN order, and the second term mainly comes from the contribution of the tensor quadrupole radiation of 2.5PN order.
Although the scalar monopole radiation is the leading term of the multipole expansion of scalar radiation, it is of the same 2.5 PN order as the quadrupole radiation.
The scalar monopole radiation carries energy but not angular momentum, because the scalar field is spin-0.
In the limit of $\epsilon_1\rightarrow 0$ and $\epsilon_2\rightarrow 0$, these results reduce to those in GR.

\section{PK parameters in GR}\label{appendix_GRPK}
In GR, the PK parameters can be related to the masses of the two bodies and to measured Keplerian parameters by the equations,
\begin{eqnarray}
\label{GRPK}
\dot{\omega}^{\gr}
&=& \Big( \frac{2\pi{Gm}}{P_b} \Big)^{\frac{2}{3}}\frac{6\pi}{P_b (1-e^2)}, \nonumber\\
\gamma^{\gr}
&=& e\frac{P_b}{2\pi}\Big( \frac{2\pi{Gm}}{P_b}\Big)^{\frac{2}{3}}\frac{m_c}{m}\Big(1+\frac{m_c}{m}\Big), \nonumber\\
r^{\gr}
&=& G{m_c},  \nonumber\\
s^{\gr}
&=& \Big(\frac{2{\pi} Gm}{{P_b}}\Big)^{\frac{2}{3}}\frac{{x_p}}{Gm_c}, \\
\dot{P}_b^{\gr}
&=& -\frac{192\pi}{5} \Big( \frac{2{\pi}Gm}{P_b} \Big)^{\frac{5}{3}}\frac{\mu}{m} F(e), \nonumber\\
\dot{e}^{\gr}
&=& -\frac{608\pi}{15} \Big( \frac{2{\pi}Gm}{P_b} \Big)^{\frac{5}{3}}
\frac{{\mu} e (1 + \frac{121}{304}e^2)}{m P_b (1 - e^2)^{\frac{5}{2}}}, \nonumber\\
\dot{a}^{\gr}
&=& -\frac{64}{5}\Big( \frac{2{\pi}Gm}{P_b} \Big)^{2}\frac{\mu}{m}F(e), \nonumber
\end{eqnarray}
with
\begin{eqnarray}
\label{F_e}
F(e) \equiv (1-e^2)^{-\frac{7}{2}}\Big(1+\frac{73e^2}{24}+\frac{37e^4}{96}\Big),
\end{eqnarray}
where $m\equiv m_p+m_c$ is the total mass, and $m_p$, $m_c$, $P_b$, $e$ and $x_p$ are the pulsar mass, companion mass, orbital period, orbital eccentricity and projected semi-major axis of the pulsar orbit, respectively.


\bibliography{fR_PPK}

\end{document}